\documentclass{aastex}
\usepackage{emulateapj5,apjfonts,psfig}

\makeatletter

\newenvironment{figurehere}
  {\def\@captype{figure}}
  {}
\makeatother

\def\cm{{\rm\thinspace cm}}

\def\erg{{\rm\thinspace erg}}

\def\keV{{\rm\thinspace keV}}
\def\km{{\rm\thinspace km}}

\def\Lsun{\hbox{$\rm\thinspace L_{\odot}$}}

\def\Msun{\hbox{$\rm\thinspace M_{\odot}$}}

\def\ph{{\rm\thinspace ph}}
\def\s{{\rm\thinspace s}}

\def\cmps{\hbox{$\cm\s^{-1}\,$}}

\def\ergpcmsqps{\hbox{$\erg\cm^{-2}\s^{-1}\,$}}

\def\ergps{\hbox{$\erg\s^{-1}\,$}}

\def\kmps{\hbox{$\km\s^{-1}\,$}}

\def\pcmsq{\hbox{$\cm^{-2}\,$}}

\def\phpcmsqps{\hbox{$\ph\cm^{-2}\s^{-1}\,$}}

\def\psqcm{\hbox{$\cm^{-2}\,$}}

\def\powerlawfluxat1kev{\hbox{$\ph\cm^{-2}\s^{-1}\keV^{-1}$}}

\def\lapp{\ifmmode\stackrel{<}{_{\sim}}\else$\stackrel{<}{_{\sim}}$\fi}

\def\gapp{\ifmmode\stackrel{>}{_{\sim}}\else$\stackrel{>}{_{\sim}}$\fi}
\def\spose#1{\hbox to 0pt{#1\hss}}

\def\approxlt{\mathrel{\spose{\lower 3pt\hbox{$\sim$}}
        \raise 2.0pt\hbox{$<$}}}
\def\approxgt{\mathrel{\spose{\lower 3pt\hbox{$\sim$}}
        \raise 2.0pt\hbox{$>$}}}

\def\lapp{\ifmmode\stackrel{<}{_{\sim}}\else$\stackrel{<}{_{\sim}}$\fi}
\def\gapp{\ifmmode\stackrel{>}{_{\sim}}\else$\stackrel{>}{_{\sim}}$\fi}

\def\mcg6{MCG$-$6-30-15}
\def\mr2251{MRC~2251-178}
\def\ngc2110{NGC~2110}
\def\iras13349{IRAS~13349+2438}
\def\iras18325{IRAS~18325--5926}
\def\grs1915{GRS~1915+105}
\def\xtej1748{XTE~J1748-288}
\def\chandra{{\it Chandra }}

\def\rxte{{\it RXTE }}
\def\asca{{\it ASCA }}

\def\xtegammamcg6{$\Gamma=1.9$}

\def\fe25{Fe~{\sc xxv}}
\def\fe26{Fe~{\sc xxvi}}
\def\Ne9{Ne~{\sc ix }}
\def\ne10{Ne~{\sc x }}
\def\mg11{Mg~{\sc xi }}
\def\si13{Si~{\sc xiii }}

\def\apj{ApJ}

\def\apj{ApJ}
\def\apjs{ApJS}
\def\aap{A\&A}

\def\c2{{\sc C~ii}}
\def\c3{{\sc C~iii}]}
\def\c4{{\sc C~iv}}
\def\n5{{\sc N~v}}
\def\o3{[{\sc O~iii}]}

\def\si4{Si~{\sc iv}}
\def\fe25{Fe~{\sc xxv}}
\def\fe26{Fe~{\sc xxvi}}
\def\mg2{Mg~{\sc ii}}
\def\Msun{\ifmmode M_{\odot} \else $M_{\odot}$\fi}
\def\Lsun{\ifmmode L_{\odot} \else $L_{\odot}$\fi}

\slugcomment{Accepted for publication in The Astrophysical Journal, 2001 November 6}

\shorttitle{Chandra-RXTE observations of \grs1915}

\shortauthors{LEE et al.}

\begin{document}

\title{High resolution Chandra HETG and RXTE observations of \grs1915 : \\ 
A hot disk atmosphere \& cold gas enriched in Iron and Silicon }
\author{
Julia C. Lee\altaffilmark{1},
Christopher S. Reynolds\altaffilmark{2,3},
Ronald Remillard \altaffilmark{1}, 
Norbert S. Schulz\altaffilmark{1}, \\
Eric G. Blackman\altaffilmark{4},
Andrew C. Fabian \altaffilmark{5},
 }
\altaffiltext{1}{Massachusetts Institute of Technology, Center for Space Research, 77 Massachusetts Ave. NE80, Cambridge, MA 02139.}
\altaffiltext{2}{Dept. of Astronomy, University of Maryland, College Park  MD 20742}
\altaffiltext{3}{JILA - U. of Colorado, Campus Box 440, Boulder CO 80309; Hubble Research Fellow }
\altaffiltext{4}{Dept. of Physics \& Astronomy, University of Rochester, Rochester, NY 14627}
\altaffiltext{5}{Institute of Astronomy, Cambridge University, Madingley Rd., Cambridge CB2 0HA  U.K.}

\begin{abstract}
The time-averaged 30~ks {\it Chandra} AO1 HETGS observation of the 
micro-quasar GRS~1915+105 in the low hard state reveals for the first time in
this source neutral K absorption edges from iron, silicon, magnesium, and sulphur.
Ionized resonance $(r)$ absorption from H- and He- like species of Fe~({\sc xxv, xxvi}) and
possibly Ca~{\sc xx} are also seen as well as possible emission from neutral
Fe~K$\alpha$ and ionized Fe~{\sc xxv} (forbidden, or the resonance emission component
of a P-Cygni profile). We report the tentative detection of the first astrophysical
signature of XAFS in the photoelectric edge of Si (and possibly Fe and Mg), 
attributed to material in grains.
The large column densities measured from the 
neutral edges reveal anomalous Si and Fe abundances, and illustrate the 
importance of high resolution X-ray measurements for assessing material
which surrounds bright X-ray sources, especially if depletion onto grains
plays a prominent role. Scenarios for which the anomalous abundances can 
be attributed to surrounding  cold material associated with GRS~1915+105
and/or that the enrichment may signify either a highly unusual supernova/hypernova,
or external supernova activity local to the binary are discussed. 
We attribute the ionized features to a hot disk, disk-wind, or corona environment. 
Based on H- and He-like Fe~({\sc xxv,~xxvi}), we estimate
constraints on the ionization parameter (log~$\xi \approxgt 4.15$), 
temperature (T~$> 2.4 \times 10^6$~K), and hydrogen
equivalent number density ($\,n \approxgt \, 10^{12} \, \rm cm^{-3}$) for this region.
Variability studies with the simultaneous \rxte data show that the light curve
count rate tracks the changes in the disk blackbody 
as well as the power-law flux, with the most significant variations observed in the 
former.  Similar studies of the \chandra spectra show spectral changes
which also track the behavior of the light curve, and may point to changes in
both the ionizing flux and density of the absorber.
A 3.69~Hz QPO and weak first harmonic is seen in the \rxte data.

\end{abstract}


\section{Introduction}
The Galactic X-ray transient GRS~1915+105 is an extremely energetic
X-ray binary that has been studied extensively in multiple wavebands
which include radio, infrared and X-rays (e.g. see review by Mirabel
\& Rodr\'iguez 1999).  Since its discovery by {\it Granat/WATCH} in
1992 (Castro-Tirado et al. 1992), it has shown repeated flares
separated by intervals in a ``low-hard'' state.    In the X-ray band,
the overall variability of \grs1915 is extremely complex, with a
luminosity which varies between a few $\rm \times 10^{37} - \, few \times 10^{39}
\ergps$,  making it one of the most energetic objects known in the
Galaxy.   Many of these variations can be understood in terms of
instabilities that switch the origin of radiant energy between two
components: the accretion disk and the X-ray power-law (Muno, Morgan,
\& Remillard 1999).
Recent observations in the near-infrared H \& K band with the VLT reveal
the donor to be a K-M~III giant (Greiner et al,. 2001a) with a 
33~day binary period and  a radial velocity curve interpreted to 
suggest a compact object of $\sim 14 M_{\odot}$ (Greiner et al, 2001b).
This indicates that \grs1915 belongs to
the class of dynamically established black hole binaries.

One of the most remarkable aspects of this system is that X-ray flares
are sometimes followed by superluminal ejection events seen in radio
images (Mirabel \& Rodr\'{i}guez 1994; Fender et al. 1999) --- the
inferred velocity of the knots in the bipolar jets is $v\sim 0.95c$
(i.e. a Lorentz factor of 3).  Only a few sources in the Galaxy
(e.g. GRO~1655--40 and XTE J1748-288) display such violent ejections,
while other sources exhibit intervals of more
constant radio emission associated with a steady jet (Mirabel \&
Rodr\'{i}guez 1999).  Both types of jet sources
qualitatively resemble radio-loud active galactic nuclei (AGN) and are
often called `micro-quasars'.

A key piece to the puzzle for understanding the physical environment of 
\grs1915 is X-ray spectroscopy.  \asca observations of \grs1915 have revealed 
striking structure within the (medium resolution) X-ray spectra.
Based on observations in 1994--1995, Kotani et al. (2000) identified 
resonant absorption lines of Ca~{\sc xx}~K$\alpha$, Fe~({\sc xxv, xxvi})~K$\alpha$,
as well as blends of absorption lines of Ni~{\sc xxvii}~K$\alpha$ + 
Fe~{\sc xxv}~K$\beta$ and Ni~{\sc xxviii}~K$\alpha$ + Fe~{\sc xxvi}~K$\beta$.
However, \asca is fundamentally limited in the extent to which
it can clearly resolve absorption lines or detect emission components.
Sufficient spectral resolution and energy coverage (such as presently possible
with the \chandra high energy transmission grating) is particularly important for obtaining direct information
on the line--of--sight velocity and velocity dispersion of the material.
The absolute and relative equivalent widths of the lines can be used to 
determine the column density and ionization parameter of the accretion disk
environment, and any disk wind. In this paper, we present primarily the 
prominent spectral features of
the  time averaged X-ray spectrum of \grs1915 obtained with the  high
resolution gratings on-board {\it Chandra}.  We assess the chemical 
abundances based on direct measurements of neutral~K edges of Fe, S, Si, and
Mg (\S\ref{sec-absedges}).   Measurements of ionized Fe features are used to assess the 
conditions of the accretion disk corona (\S\ref{sec-fe}).  
Variability is discussed 
in the context of \chandra and \rxte observations (\S\ref{sec-var}). 
 Preliminary results were presented in Lee~et~al.~(2001a).

\section{Observations}

\grs1915 was observed with the {\it Chandra} High Energy Transmission
Grating (HETGS; Canizares et al., 2001, in preparation) on 2000 April 24
beginning at 01:35:49 (MJD: 51658.06654), with a total integration time 
$\sim$~31.4~ks.    This corresponds to orbital phase~$\sim 0.8$ 
according to the ephemeris quoted by Greiner (2001b), where 
phase~zero (2000 May 02) is the time consistent with the 
 blue-to-red crossing in the radial velocity.
Simultaneous \rxte
observations were performed  with both the Proportional Counter Array
(PCA, Jahoda et al., 1996) and High-Energy X-ray Timing Experiment
(HEXTE, Rothschild et al., 1998) instruments.

Observations with the \rxte All-Sky Monitor over the time interval of
2000 April 21-28 indicate a relatively steady X-ray flux (16\% rms
variations) and a hard spectrum, with an average intensity of 0.4 Crab
at 2-12 keV.  Radio observations of GRS~1915+105 with the Greenbank
Interferometer
\footnote{http://www.gb.nrao.edu/fgdocs/gbi/gbint.html} on 2000 April
24.54 indicate a flux of $20 \pm 4$ mJy at 2.25 GHz. The combined
radio and X-ray  properties place GRS~1915+105 in the "low-hard" state
that is associated with the presence of a steady jet (Dhawan, Mirabel,
\& Rodriguez 2000; Muno, Remillard, \& Morgan 2001).

Both the HETGS and the \rxte PCA instruments detect variations  from
GRS~1915+105 in the form of a smooth, 15\% dip in the  count rates
that occurs midway through the observations (Fig.~\ref{fig-variability}a,
top 2 panels).  The spectral variations
are subtle, and so we first consider the average spectral features and
then examine the dip in greater detail in \S\ref{sec-var}.

\subsection{The Chandra data \& reduction}  \label{sec-chandradata}
The \chandra HETGS is made up of the Medium Energy Grating (MEG) with
a 2.5-31~\AA\, (0.4-5~keV) bandpass, and High Energy Grating (HEG)
with 1.2$-$14~\AA\,(0.9-10~keV) bandpass.  The peak resolving power
(E/$\Delta$E) for the MEG and HEG is $\sim$1000.  The resolution of
the higher orders improves by a factor of $n$ for the $n$th order, but
the spectral bandpass and efficiency are reduced accordingly.

We observed \grs1915 during a persistent phase of relatively low flux.
Despite this, the observation suffered from photon pile-up.  This is
the phenomenon by which multiple photons fall on a single pixel during
the same readout period. As a result, the on-board event detection
algorithm interprets this as a single high energy photon event,  and
sums the multiple photon events to make a single pulse height.  Pileup
has the effect of depleting the photon spectrum, thereby 
decreasing the photon counting rate.
Corrections for this effect will be discussed in \S\ref{sec-continuum}.

Due to severe telemetry and photon pileup problems imposed by the
large count rate of \grs1915, the observation was performed using the
{\it graded}
\footnote{See \chandra Proposer's Observatory Guide at
http://asc.harvard.edu} telemetry mode, and a subarray which reduced
CCD frametime by a factor of $\sim 2$.  We additionally used a
`spatial window' to block out the 0th order image (this would have
been completely piled-up, even during periods of low flux for the
energetics of \grs1915).  However, because the 0th order position is
crucial for defining an accurate wavelength  scale, we calculate its
position for our observation  by finding the intersection of the
dispersed MEG--HEG spectra with the 0th order readout trace.  The
accuracy of such a technique for the determination of the 0th order
position is $\sim$ 0.2$-$0.3 pixels, which translates respectively to
a wavelength accuracy of 0.002~\AA \, and 0.004~\AA\, for the HEG and
MEG 1st orders.


\subsection{The RXTE data \& reduction}
We have analyzed the \rxte observations of 2000 April 24 that overlap
the \chandra exposure of \grs1915.  At high time resolution, the  PCA
telemetry modes yielded 2~ms resolution for two energy channels that
span 2--9~keV.   We combined these channels and computed a power
density spectrum, corrected for  instrument deadtime and counting
statistics (see Morgan, Remillard, \& Greiner 1997).  We then
normalized the power spectrum to units of (rms
fluctuations/mean)$^2$~Hz$^{-1}$  in the source, applying a small
correction for the contribution of the diffuse X-ray background
between 2--9~keV.  The results show a QPO and a broadband power
continuum typical of the low-hard state  in \grs1915 (e.g., Muno et al.,
2001).  The strongest QPO feature is at  3.69~Hz with an integrated
amplitude of 8.5\% and a weak first harmonic.

PCA and HEXTE spectra were derived from standard data modes that
produce 128--channel PCA spectra (full bandwidth) every 16~s, and
64--channel HEXTE spectra  (full bandwidth) at the instrument nodding
period of 32~s.  There were 3 PCUs in operation (units 0, 2, and 3)
during the PCA observations, and our spectral analyses focus on
results from PCUs 0 and 2 because of the superior performance of the
associated calibration files ({\sc ftools 5.6}), as judged from the
analysis of PCA observations of the Crab Nebula during late 2000 and
early 2001.

\section{The Spectrum}   \label{sec-continuum}
The best fit continuum model which describes the \rxte data is a
disk-blackbody plus a power-law component each modified by a hydrogen
column of $5 \times 10^{22} \psqcm$.  The best fit color temperature
of the disk is 1.4~keV ($\sim 1.6 \times 10^7$~K) with a normalization of
$N_{\rm bb} = 59$.  (We note that the temperature deduced from this black
body fit and that in the subsequent calculation of \S\ref{sec-fe} based on
the ratio of Fe~{\sc xxv}~:~Fe~{\sc xxvi} are not necessarily in conflict
given that the {\sc diskbb} model is expected to overestimate the 
temperature by as much as a factor of 1.7 [Shimura \& Takahara 1995],
and the value given in \S\ref{sec-fe} is a lower limit;  however,
the temperature measurements here and in \S\ref{sec-fe} can also 
be probing independent components of the gas.)  The average
power-law component exhibits a photon index of $\Gamma \sim 2.4$ and a
normalization of the power-law at 1~keV is 11.4~$\powerlawfluxat1kev$.
The corresponding apparent (since there are losses due to low energy 
absorption) total 2--25~keV flux is $1.89 \times 10^{-8}$~\ergpcmsqps.
The implied (unabsorbed) bolometric fluxes are respectively 
$\sim 5.3 \times 10^{-9}$~\ergpcmsqps and $\sim 3.4 \times 10^{-8}$~\ergpcmsqps
for the disk-blackbody and power-law (integrated from 1~keV to $\infty$)
components, indicating that the power-law component was the
dominant emission mechanism during these observations.  
The absolute absorption corrected luminosity corresponding to the epoch
of our observation is lower limit of $L_{bol} \approx L_X \approx  6.4 \times 10^{38}$ \ergps at
12.5~kpc.
We also assess the higher energy data 
afforded by \rxte HEXTE observations, and find that it is 
best fit with a broken power-law with the cutoff $E_{\rm break}$ = 30~keV.
The photon index below 30~keV is $\Gamma \sim 2.4$ (consistent 
with PCA measurements), and a steeper $\Gamma \sim 2.9$ spectrum
above 30~keV.
Since \rxte is better suited for
constraining the continuum shape due to  its wide band coverage, we
apply this model to the \chandra data.  \\

\vspace{0.2in}
\begin{figurehere}
\vspace{0.2in}
\centerline{\psfig{file=grs1915_fig1.ps,angle=270,width=8.5truecm,height=8.5truecm}}
\figcaption{Best-fit \chandra continuum (black) based on best-fit \rxte continuum model over-plotted on the (a)
non-pileup corrected \chandra HEG 1st order (blue), MEG 1st order (red), and MEG 3rd (red) order
and (b) pileup corrected HEG 1st order (blue).
\label{fig-continuum}}
\end{figurehere}
\vspace{0.2in}

The overall shape of the continuum discussed above describes the
\chandra data well (with the exception that the best fit power-law 
and power-law normalization are respectively $\Gamma \sim 2.1$ and 
5.3~$\powerlawfluxat1kev$ for fits to the \chandra data), especially given the problems with pileup,
which is responsible for 
$\sim$~40\% of the  60\% normalization difference between the \rxte
and \chandra data.  (Pileup will affect different parts of the 
spectrum by different amounts.)
To illustrate, Fig.~\ref{fig-continuum}a shows the
absorbed disk-blackbody plus power-law model over-plotted on the HEG
1st and MEG 1st \& 3rd order data.  Spectral  deviations from the
model in the MEG/HEG 1st order spectra at $\approx$~2-5~\AA\, is due
to pileup as a result of the high count rate of the source in this
passband; instrumental pileup severely affects the 1st orders between
$\approx$~6.5-9~\AA\, since this is where the effective area of the
HETGS peaks.  The MEG 3rd order spectrum at $<$~2~\AA\, is affected by
both first order photon pileup at $\sim$~5~\AA\, (i.e. {MEG 3rd order
pileup = $\lambda$/order = 5\AA/3}), and 0th order scattering.  
We extract a separate set of HEG and MEG 1st order spectra to be used 
with the HETG pileup model as implemented in 
{\sc isis}\footnote{http://space.mit.edu/CXC/ISIS/}
(see Davis 2001, Houck \& DeNicola 2000).   These data
are extracted with a coarser (respectively for the HEG and
MEG 1st orders $\sim$0.017\AA\, and 0.033\AA\,) binning to 
reflect the size of an ACIS event detection cell.  
The pileup corrected data shows a reduction from 60\% to a few~\% in 
the overall normalization offset between \rxte and \chandra (Fig.~\ref{fig-continuum}b).
We note that the broadband
\rxte PCA flux is typically $\sim 15-25$\% higher when compared
against other X-ray missions (e.g. \asca, HEXTE). \\

\centerline{\psfig{file=grs1915_fig2.ps,angle=0,width=8.5truecm,height=8.5truecm}}
\figcaption{Photoelectric K-shell edges of S, Si and Mg.  Overplotted are the respective
best-fit models described in \S\ref{fig-edge}.
Note the prominence of the Si edge, and possible XAFS structure.
(The Fe~K edge is shown in Fig.~\ref{fig-feregion}.) \\
\label{fig-edge}}

\section{Photoelectric edges} \label{sec-absedges}
We find evidence for prominent absorption edges due to K-shell Fe, S, Si, and
Mg (Table~1, Figs.~\ref{fig-edge}~\&~\ref{fig-feregion}).
Because the spectrum of \grs1915 is complicated by ionized resonance features
(e.g. Fig.~\ref{fig-feregion}), and possibly resonance structure
due to X-ray absorption fine structure (XAFS),
we derive $\tau$ by fitting 
a linear function to pieces of the data bracketing $\sim 0.1$~\AA\, to either 
side of a given edge, and extrapolate the fit to the expected edge energy 
(see Fig.~\ref{fig-edge}).  (XAFS in the spectrum of \grs1915 are discussed 
later in this section.)
Additionally, since pileup will also affect the depth of an edge 
by reducing the jump in the discontinuity at the edge,  we use
the pileup corrected 1st order HEG to assess the optical depth of Fe, S, \& Si, and MEG to assess Mg.
The column densities (shown in Table~1) are calculated  using the
value for the K-shell photoionization cross section of the relevant
species (~\footnote{http://www-cxro.lbl.gov/optical\_constants/asf.html}~Henke et al., 1993), 
and evaluated for solar and ISM abundances (Wilms, Allen, McCray 2000).  Given the
possibility for emission and absorption features and edge structure
which we have not accounted for, a conservative estimate for the
systematic uncertainties in $\tau$ is calculated using the error in the 
counts and best fit continuum level on either side of the edge to obtain
a fractional uncertainty.

\vbox{
\footnotesize
\begin{center}

\begin{tabular}{lcccc}
\multicolumn{5}{c}{\sc Table 1} \\
\multicolumn{5}{c}{\sc Prominent Neutral K Edges of the Time-averaged data} \\
\hline
\hline
\multicolumn{5}{c}{\sc Photoelectric Edges} \\
{\em \rm Species} & {\em \rm $\lambda_{0}$ }& {\em \rm $\tau$}  & {$^a \rm N_{\rm H}$ (solar)} & {$^b \rm N_{\rm H}$ (ISM)} \\
&   {(\AA)} & & {$(10^{22} \pcmsq$)} & {$(10^{22} \pcmsq$)}   \\
\hline
Mg K  & 9.52 & $ 0.29 \pm 0.03 $ & $3.1^{+0.3}_{-0.3}$  & $4.7^{+0.5}_{-0.5} $ \\
Si K  & 6.74 & $ 0.47  \pm 0.01$ & $ 8.4^{+0.1}_{-0.2}$  & $16.1^{+0.2}_{-0.3}$ \\
S K & 5.02 & $  0.07^{+0.002}_{-0.01} $ & $3.2^{+0.1}_{-0.6} $  & $4.8^{+0.2}_{-0.9}$\\
Fe K & 1.74 & $ 0.11  \pm 0.02$ & $ 9.3^{+1.6}_{-1.3}$ & $10.9^{+1.9}_{-1.5}$ \\

\hline
\end{tabular}

\parbox{3.2in}{
\vspace{0.1in}
\small\baselineskip 9pt
\footnotesize
\indent
$\rm ^{a, b}$~The total hydrogen column assuming (a) solar and (b) ISM abundances.
The expected atomic ($N_{\rm H}$) and molecular ($N_{\rm H2}$) hydrogen along the line of sight is
respectively $\approxgt 1.76 \times 10^{22} \pcmsq$ and $\sim 7.43 \times 10^{21} 
\pcmsq$. The MEG summed 1st order data is 
used to assess the Mg edge; all other edges were fit using the HEG summed 1st 
order data. \\
}
\label{tab-edge}
\end{center}
\normalsize
\centerline{}
}

We compare our results (below) with constraints on the
Galactic~\footnote{http://heasarc.gsfc.nasa.gov/cgi-bin/Tools/w3nh/w3nh.pl}~$N_{\rm
H_G}  \approxgt 1.8 \times 10^{22} \pcmsq$ (determined from 21~cm
emission from atomic hydrogen along the line of sight, Dickey \&
Lockman, 1990).  Despite the corrections for optical depth effects,
this value is regarded as a lower limit 
because of its reduced sensitivity to both cold and ionized H, 
as well as H$_2$.  Dame et al. (2001)  report additional molecular hydrogen 
(based on CO maps) with $N_{\rm H2} \sim 7.43 \times 10^{21}  \psqcm$ to the 
nearest $l,b$ towards \grs1915.  Therefore, the line of sight column can be 
$\approxgt 3 \times 10^{22} \, \rm cm^{-2}$ with contributions from both 
atomic and molecular hydrogen.  (Chapuis and Corbel 2001, in preparation,
 find the total [atomic plus molecular] line of sight column density 
toward \grs1915 from millimeter and HI observations to be $3.3 \pm 0.2 \times 10^{22} \psqcm$;
private communication S. Corbel.)

Using the HETG absorption edges for S and Mg, we derive $N_{\rm H}$
values $\sim 3 \times 10^{22}$ cm$^{-2}$, assuming solar abundances, and
$\sim 5 \times 10^{22}$ cm$^{-2}$, assuming ISM abundances.
These values are consistent with the $N_{\rm H}$ value derived from
low energy continuum absorption.  However, the
$N_{\rm H}$ derived from Si \& Fe (for solar
abundances) are greater by $\approxgt 5 \times 10^{22} \pcmsq$ (and
much greater for ISM abundances) when compared with what is expected
from the line of sight column density (Table~1).
We note that earlier reports by Lee
et al. (2001a) for $N_{\rm H}$ were based on data that had not been corrected
for pileup.

We point out the exciting possibility that astrophysical XAFS
are present in the spectrum of \grs1915, most noticeable
in Si, but may also be present at the Mg and Fe edges.  
XAFS is a general term for
both  EXAFS (extended X-ray absorption fine structure) and
XANES (X-ray absorption near edge structure), with the distinction
based on single (EXAFS) versus multiple (XANES) scatterings
of the photoelectrons from the immediate neighboring atoms (e.g.
Sayers, Stern, \& Lytle 1971, Woo, Forrey, \& Cho 1997, and references therein).   
Such features appear as wave--like modulations, with the 
former manifest in structure at $\approxgt 50$~eV from the absorption edge,
and the latter in the region between the edge and EXAFS region.
A qualitative comparison of Si in the detector ($\rm SiO_2$ and polysilicon)
shows that the modulations of the Si EXAFS in our observation are much stronger than
that expected from the detectors, and therefore may be features 
which are associated with the X-ray source.
We note further that the detector Si region has been carefully
calibrated against ground based measurements at synchrotron beamlines
(e.g. Prigozhin et al., 1998; also Mori et al., 2001 for further studies of 
absorption edges in front illuminated CCDs).   Therefore, we believe that 
a definitive claim of XAFS in \grs1915 is hampered 
more by the lack of adequate statistics than calibration errors in 
the quantum efficiency of the \chandra CCDs. (The calibration near the Si edge is good 
to $\sim 3\%$ over 0.02~\AA \,bins -- private communication, H. Marshall.)  The possible EXAFS 
at the Si edge are noted in the instrument-corrected spectrum 
shown in Fig.~\ref{fig-edge}.   
An investigation of several slightly less absorbed X-ray binaries
which have good statistics and for which the instrumental response was 
applied (Schulz et al., 2000) do not
show XAFS such as that seen here, and gives us further confidence that the XAFS seen in the \chandra
spectrum of \grs1915 is associated with an astrophysical phenomenon.
At present, we merely point out this exciting possibility --
an in-depth treatment of XAFS is beyond the scope of this paper, and 
will be addressed in a subsequent paper which will investigate the
presence of XAFS in highly absorbed bright X-ray binaries.
 
We further verify that the Si edge and possible structure 
is intrinsic to the source by comparing the measured edge depth 
against the  contribution from the
\chandra CCDs.  The discontinuity in the  effective area at the
detector Si edge is  $\sim 20-30$\%, and verified to be accurately 
\footnote{CXC Proposers' Observatory Guide, December 2000, Version 3, p.190}~modeled 
to $\approxlt 10$\%
but as stated previously can be good to $\sim$~3\% over 0.02~\AA \,bins;
the latter is particularly important for the assurance that the residuals
which we attribute to XAFS at the Si edge are not likely to be an instrumental
feature.  This is
compared with the $>$ factor 2 discontinuity measured from the  HEG 1st order
count spectra.  We find further confidence that the Si edge is not a
detector feature from the MEG 3rd order spectrum, where the  edge
remains prominent in the data, but for which there  is a minute
contribution from the detector.  (The spectra for the relevant orders
are the combined plus and minus sides;  for the MEG 3rd order, the Si
edge seen in our spectrum for one of the sides fall on the backside
(S1) chip where there is no/little contribution from the  detector Si
edge.  -- See e.g. the effective area curves shown in Fig.~2 of Lee et al., 2001a)

\begin{table*}[t]
\begin{center}
\begin{tabular}{lcccccc}
\multicolumn{7}{c}{\sc Prominent Absorption \& Emission Features in the Time-averaged data} \\
\hline
\hline
{\em \rm Species} & {$^a$ \em \rm $\lambda$} & {\em \rm $^b \lambda_{mea}$} & {$^c \Delta \lambda$ } & {$^d \sigma$ } & {\rm $^e$ Flux} & {$^e W_{\rm Fe}$} \\
&  {\em \rm (\AA)}  & {\em \rm (\AA)}  & {($\kmps$)} & {($\kmps$)} & {($10^{-3}\, \phpcmsqps$)} & {\em \rm (m\AA)}\\
\hline
Fe~{\sc xxvi}~Ly~$\alpha$$^\ast$  & 1.7798 & $ 1.7752 \pm 0.0024 $ & $ -770 \pm  400 $ & $ 578 \pm 400$ & $-1.72 \pm 0.41 $ & 4.7  \\
Fe~{\sc xxv}~Ly~$\alpha$$^{\ast 1}$  & 1.8505 & $  1.8547 \pm  0.0062 $ & $ +680 \pm  1000 $  & \small{= Fe~{\sc xxvi}~Ly~$\alpha$} & $ -0.61 \pm 0.38  $  & 1.7 \\
Fe~{\sc xxv}~Ly~$\alpha$  & 1.8505 & $  1.8540 \pm 0.0070 $ & $  +570 \pm 1140  $  & $ 499 \pm 728$ & $ -0.58 \pm 0.37 $  & 1.6 \\
Fe~{\sc xxv} $^{\dagger 1}$ & 1.8682 & $ 1.8676 \pm 0.0027  $ & $  -100 \pm 430 $ & \small{= Fe~{\sc xxvi}~Ly~$\alpha$}  & $ 0.98 \pm 0.36$ & 2.7 \\
Fe~{\sc xxv}~$^{\dagger}$ & 1.8682 & $ 1.8685 \pm 0.0028 $ & $ +50 \pm 450 $ & $ 164 \pm 346 $ & $1.00 \pm 0.36$ & 2.7 \\
Fe K$\alpha$  & 1.9361 & $ 1.9350 \pm 0.0094 $ & $ -170 \pm 1457 $  & $ 380 \pm 670 $ & $ -0.71 \pm 0.35 $ & 2.0  \\
Un-IDed  & $--$ & $1.7315 \pm 0.0016 $ & $ -- $  & $ 177 \pm 165 $ & $\approxlt 1.96 \pm 0.51$ & $\approxlt  6$\\
Ca~{\sc xx}~Ly~$\alpha$$^{\ast 2}$ & 3.0203 &  $ 3.0117 \pm 0.0029$ & $ -850 \pm 290$ & $205 \pm 135$ &  $> 0.34 \pm 0.17$  &  $> 1.2$ \\

\hline
\end{tabular}
\end{center}
\setcounter{table}{1}
\noindent
\caption{\noindent  \rm
$\rm ^\ast$Absorption feature; otherwise emission.
$\rm ^\dagger$Line is identified as either the Fe~{\sc xxv}~($f$) emission or the emission portion of the Fe~{\sc xxv}~($r$) P-Cygni.
$\rm ^{1}$The line width is tied to that of  Fe~{\sc xxvi} absorption.
$\rm ^{2}$Lower limit values quoted since region is heavily piled-up.
$\rm ^a$~Rest wavelengths of identified lines from Verner et al., 1995
$\rm ^b$~Best measured centroid wavelengths (\AA).
$\rm ^c$~Amount of shift of measured wavelength from rest (- for blueshift, + for redshift)
$\rm ^d$~`True' velocity width (km/s).  The instrumental line spread function has already been subtracted out. 
$\rm ^e$~Flux in the line.
$\rm ^f$~Equivalent width. }
\label{tab-reslines}
\end{table*}

\section{The iron complex - a hot disk atmosphere ?}  \label{sec-fe}

In addition to neutral edges, and possible neutral (Fe K$\alpha$)
emission,  the spectrum of \grs1915 also exhibits a strong Fe~{\sc xxvi} 
resonant absorption feature (Fig.~\ref{fig-feregion}).   A weak Fe~{\sc xxv} resonant absorption
may also be present.    We detect a $\sim$2.7$\sigma$ emission feature 
at $\sim 1.87$\AA\, which can be either the  Fe~{\sc xxv} $1s^2 \, ^1S_0\,  -\,  1s2s\, ^3S_1$
(forbidden) emission, or the emission component of a P-Cygni resonance line of 
 Fe~{\sc xxv} (similar to what is seen in Cir~X-1 - Schulz \& Brandt, 2001, in preparation).
Because the Fe region is $< 5$\% piled-up, we use the original HEG combined plus and minus
1st order data binned to 0.0075~\AA\, to assess the features detailed in Table~\ref{tab-reslines}.
The $\sim 0.002$~\AA\, wavelength errors are consistent with the expected accuracy
for the HEG data (e.g. \S\ref{sec-chandradata}), and present measurements 
show consistency with a zero velocity shift.
(The \asca spectrum of the microquasar GRO~1655-40 (Ueda et
al., 1998) also shows no evidence for blueshifted lines.)
The resolution of the \chandra HEG is $\sim 1600 \kmps$ FWHM ($\sim$10~m\AA\,) at the iron energies. 
The velocity widths of Table~\ref{tab-reslines} indicate that 
the lines are either not resolved or only marginally so.  The larger breadth
to the line width of the Fe~{\sc xxvi} absorption when compared with the 
$\lambda \sim 1.87$\AA\ Fe~{\sc xxv}~ emission (if the forbidden line) may indicate that the region responsible for the
Fe~{\sc xxv} emission may only be part of the region responsible for the
Fe~{\sc xxvi} absorption.  Alternatively, the $\lambda \sim 1.87$\AA\ line is the ($r$)
emission component of the 
Fe~{\sc xxv} P-Cygni complex which is equally divided between absorption and emission.

Because the Fe~{\sc xxvi} and Fe~{\sc xxv} resonant absorption features are detected,
we can estimate the physical parameters of the plasma from which they 
originate based on the ratio of the Fe~{\sc xxvi}~:~Fe~{\sc xxv} column densities
($N_{\rm Fe26} : N_{\rm Fe25}$).
If the lines are resolved (e.g. Fe~{\sc xxvi}), we can 
assume that they are unsaturated and on the linear
part of the curve of growth, and therefore estimate the  Fe~{\sc xxvi}
and Fe~{\sc xxv} column densities to be  
$N_{\rm Fe26} \sim 4 \times 10^{17} \, \psqcm$
and $N_{\rm Fe25} \sim 7 \times 10^{16} \, \psqcm$ 
based on the relation

\begin{equation}
\frac{W_\lambda}{\lambda} \, = \, \frac{\pi \,e^2}{m_e \, c^2}\,  N_j \, \lambda \, f_{ij}
			     = 8.85 \times 10^{-13} \,  N_j \, \lambda \, f_{ij}	\,
\end{equation}
where $N_j$ is the column density for the relevant species, $f_{ij}$
(respectively for Fe~({\sc xxv, xxvi}) is 0.798, 0.416, from Verner 1995) is the 
oscillator strength, $W_\lambda$ is the equivalent width
of the line, and $\lambda$ is the wavelength in cm units (Spitzer 1978).  
(If instead, the features are due to a
superposition of saturated lines over a range of velocities, the
column densities quoted would be higher.)
The 1$\sigma$ confidence upper limit for the Fe~{\sc xxv} flux in Table~\ref{tab-reslines} gives 
an equivalent width $\sim 2.7$ m\AA \,, and therefore $N_{\rm Fe25} \approxlt 1.1 \times 10^{17} 
\, \psqcm$ (compared with the bestfit $N_{\rm Fe25} \sim 7 \times 10^{16} \, \psqcm$
quoted above).  We use this upper limit value of $N_{\rm Fe25}$ for our subsequent 
calculations to estimate the physical conditions (e.g. temperature, ionization parameter, etc.)
 of the hot plasma.  
Results based on calculations using the best-fit Fe~{\sc xxv} flux
and equivalent width will be noted in `[]'. 

\vspace{0.3in}
\centerline{\psfig{file=grs1915_fig3.ps,angle=270,width=8.5truecm,height=8.5truecm}}
\figcaption{Time-averaged AO1 30ks HEG 1st order data of the Fe region
binned to 0.0075~\AA show both ionized (most likely from the
accretion disk atmosphere) and neutral (from the source environment,
or along the line of sight) Fe features. Over-plotted is the best fit 
continuum (dashed blue) and identified (red) lines.  The unidentified 
line may be a shift in the edge due to XAFS.
\label{fig-feregion}}
\vspace{0.3in}

Based on the calculations of
Kallman \& Bautista (2001) for a power law ionizing spectrum of energy
index $\alpha = -1$ ($\Gamma = 2$), and photoionized plasma, the ratio
of $N_{\rm Fe26}$ to $N_{\rm Fe25}$  implies an ionization fraction of
$\approxlt 0.45$ [$\sim 0.39$]
  for Fe~{\sc xxvi} from which we estimate log~$\rm \xi \approxgt
4.15 [\, \rm erg \, cm \, s^{-1}]$ [log~$\xi \sim 4.3$],
and T~$\approxgt 2.4 \times 10^6$~K [$\sim 2.7 \times 10^6$~K]. (This is a much higher 
ionization state than that reported by Kotani et al., 2000 for their high
temperature limit.)   Given the likelihood that these
ionized features originate in close proximity to the  bright X-ray
source, we expect photoionization (rather than collisional ionization)
to be the dominant physical process, due to the high radiation field.
(Temperatures of $\approxgt 10^8$~K would be necessary for  producing
hydrogen--like iron in a collisionally ionized plasma.)  We
next correct for the ionization fraction to  obtain an equivalent
hydrogen column $N_{\rm H}$ in this region respectively  for solar and ISM
abundances to be $2.8 \times 10^{22} \psqcm$ and  $3.3 \times 10^{22}
\psqcm$.  Based on the relation $\xi = L / nR^2 \rm \,erg\,cm\,s^{-1}$
where  $N_{\rm H} = n\Delta R \equiv$ the maximum column density in this
region, we estimate $R^2 / \Delta R \sim 2 \times 10^{12}$~cm. ($R$ is
defined to be the distance from the ionizing flux to the absorber 
and $\Delta R$ the thickness of the absorber.)  If we further
postulate that the volume filling factor $\Delta R / R$ must be small
(e.g. $\approx 0.1$)  in order that $\xi$ not change over the
region, we estimate an upper limit on $R \approxlt 2 \times 10^{11}$~cm,  implying that
$\Delta R \approx 2 \times 10^{10}$~cm.  It follows that the 
hydrogen equivalent number density  $n \approxgt 2 \times 10^{12} \rm cm^{-3}$
(assuming solar abundances).   

It is plausible that the ionized absorber responsible for the H- and He-like Fe lines
at a distance $R \approxlt 10^{11} \, \rm cm$ is associated 
with material flowing out from the X-ray source (e.g. wind, see also \S\ref{sec-var}), which we can compare
with the accretion rate.
Using the relation for the ionization
parameter, we can define the (spherical) mass outflow rate
$\dot M_{flow}$ at a velocity $v \sim 10^7 \cmps$ (100~$\kmps$) to be
\begin{eqnarray}
\dot M_{flow} =  4\pi r^2 n m_p v \, (\frac{\Omega}{4\pi}) & = & 4\pi m_p v \, (\frac{L_x}{\xi})\,  (\frac{\Omega}{4\pi}) \\
 & \sim & 9.5 \times 10^{18} \, (\frac{\Omega}{4\pi}) \, \rm gm \, s^{-1}    \nonumber
\label{eq-Lk}
\end{eqnarray}
where $r$ is the characteristic radius, $\rho = nm_p$ is the density of the
absorbing material, with  
$n$ equal to the number density of
electrons in this material and $m_{\rm p}$ the proton mass, and $\Omega \equiv$~solid 
angle subtended by the outflow.  
This can be contrasted with 
\begin{equation}
\dot M_{accretion} = \frac{L_{bol}}{\eta \, c^2} \sim 7.1 \times 10^{18} \, \rm gm \, s^{-1}
\end{equation}
where the efficiency $\eta \sim 0.1$ and the lower limit on the bolometric luminosity 
$L_{bol} \sim L_x$ is used.
(from \S\ref{sec-continuum}).     
Such a comparison for $v \sim 100 \kmps$ shows that as the covering fraction (i.e. $\Omega/4\pi$)
approaches unity, the mass flux in the flow (or wind) can be comparable to that in the disk.

The high ionization state of the hot medium (even based on the lower limit log~$\xi \sim 4.15$ )
 which supports the H- and He-like 
Fe lines is too ionized to support most other species with the exception
of H-like Ar~{\sc xviii} (ionization fraction $\sim 9$\%) and H-like Ca~{\sc xx} ($\sim 16$\%),
and  Fe~{\sc xxv} ($\sim 12$\%), compared to the $\sim 45\%$ ionization fraction for Fe~{\sc xxvi}.
(These percentages will decrease with increasing ionization.)
The lower limit (since this region is severely affected by pileup) to the column density of 
Ca~{\sc xx} ($f_{ij} = 0.416$) is $N_{\rm Ca20} \approxgt 3.6 \times 10^{16} \, \psqcm$

\section{Tracking variability in the disk ?} \label{sec-var}
\begin{figure*}[t]
\includegraphics[angle=0,height=3.3in,keepaspectratio=false,width=3.2in]{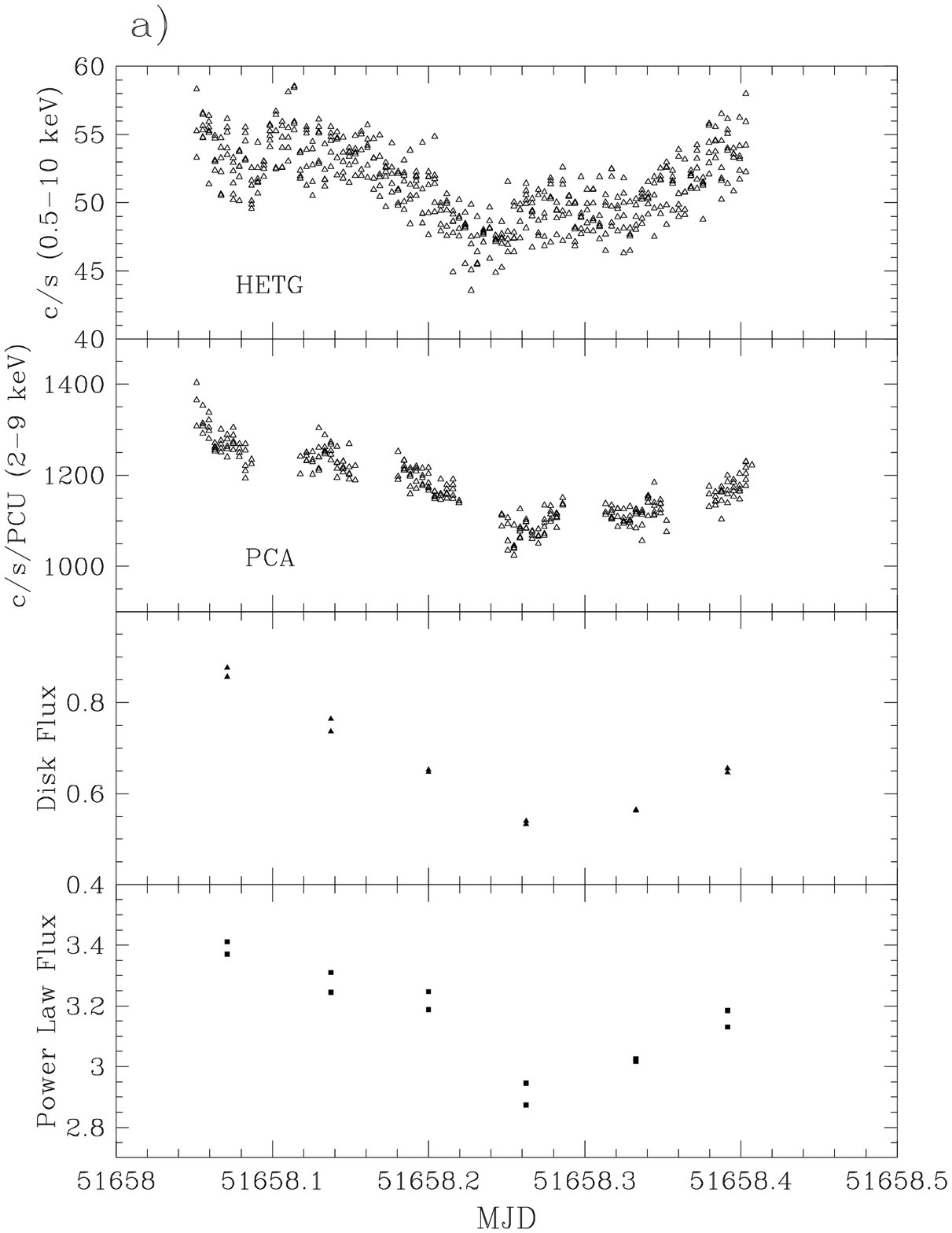}
\hspace{-0.4in}
\includegraphics[angle=0,height=3.3in,keepaspectratio=false,width=4.2in]{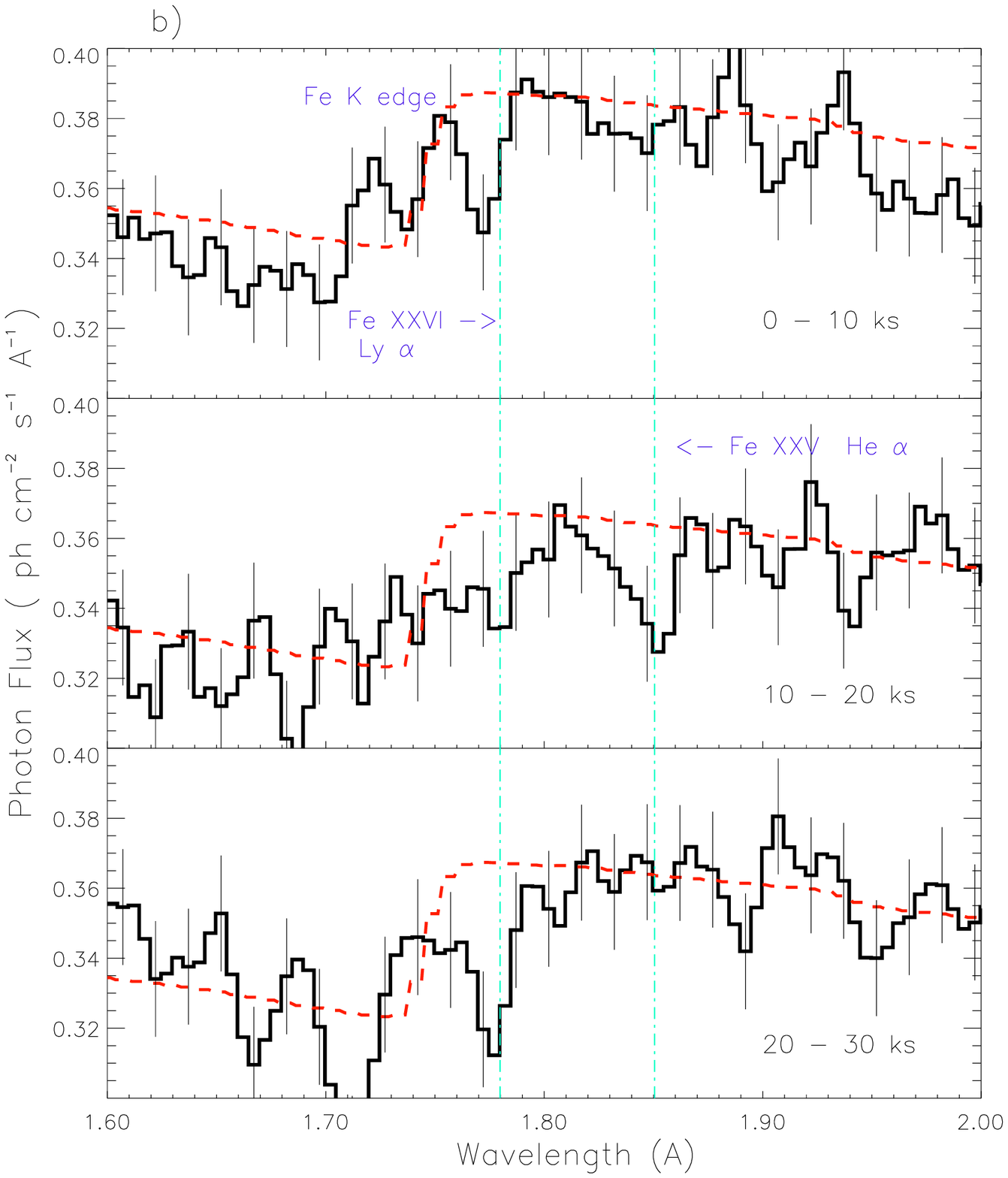}\\
\vspace{0.2in}
\caption[h]{\footnotesize (a) Simultaneous \chandra (top panel 1) and \rxte (panel 2) 
light curves during the epoch beginning 2000 April 24 at 01:35:49 (MJD: 51658.06654).  
Changes in the blackbody flux (panel 3) and power-law flux (panel 4) are coincident with changes 
in the light curve count rates.   (Panels 3 \& 4 are in units of $10^{-8}$~\ergpcmsqps.)
(b) The 30~ks \chandra spectrum divided into three 10~ks intervals depicts the evolution of the 
spectral features.  Note in particular the prominence of the Fe~{\sc xxvi} absorption at the 
beginning and end of the observation when the flux was highest, and of Fe~{\sc xxv} during the dip.
Over-plotted is the rough shape of the continuum (red-dashed) 
from the time-averaged spectra shown in Fig.~\ref{fig-feregion}. 
Error bars are representative of a statistically independent bin for 
which the data have been boxcar smoothed by 3, corresponding to $\sim 0.0075$~\AA.
}
\label{fig-variability}
\end{figure*}

The count rates from the \chandra HETGS and the \rxte PCA instruments for
GRS~1915+105 are shown in the top two panels of Fig.~\ref{fig-variability}a.
To investigate the physical processes which may be related to
the observed temporal variability,
we use the PCA spectral parameters to compute the inferred
bolometric flux from the accretion disk (Fig.~\ref{fig-variability}a, panel 3) 
and from the power law  component (1-20 keV; panel 4).
We find that the most significant variations occur
in the disk blackbody component, although there
is also a smaller dip in the power-law flux as well.

The time-sliced \chandra data shown in Fig.~\ref{fig-variability}b
suggests that evolution of the ionized line features may also be
correlated with changes to the disk flux.  In particular, the \chandra
spectra shows that the Fe~{\sc xxv} line is more prominent during the
dip between the $\sim 10-20$~ks of our observation, in contrast to the
prominence of the Fe~{\sc xxvi} absorption at the beginning and end of
the observation when the source is brighter.
 The time interval corresponding to the strongest Fe~{\sc xxv}
absorption is coincident with the dip in the
Fig.~\ref{fig-variability}a (panels 1 \& 2) light curve count rate,
as well as the notable diminution of the disk flux (panel 3), and to a lesser
extent, the power law flux (panel 4) seen in the \rxte data.   This would
support the idea that the ionized absorption features originate from
the disk.  However, we note that an $\sim$~20\% change in the flux at
the ionization discussed in \S\ref{sec-fe} implies an $\sim$ 30\%
change in the ionization fraction of Fe~{\sc xxv}, if we  assume that
$\xi$ scales with the ionizing flux, while $n$ and $R$ remain
constant.  This is not sufficient to explain the lack of Fe~{\sc xxv}
absorption, and dominance of  Fe~{\sc xxvi} absorption during the
periods when the count rate is high. An $\sim$~50\% change in $\xi$
which will reduce the ionizing fraction of Fe~{\sc xxv} by a factor of
two might be more plausible for explaining what is observed,
indicating that the ionizing flux is only part of the solution, 
and changing density the other factor.  If we consider the calculations 
in \S\ref{sec-fe} for a wind with velocity $v \sim 10^7 \, \cmps$ and distance
to the absorber $R \sim 10^{11}$~cm, it is plausible that the spectral variability 
can be attributed to a flow which can change on timescales of 10~ks such as that
seen here. This is not inconceivable since 10~ks is a very long time, corresponding
to many dynamical timescales of the relevant parts of the disk, in the
life of a source like \grs1915.  Furthermore, since this source accretes near 
its Eddington limit, one would expect the mass flow rate and wind density 
to be strong functions of the luminosity, such that the actual structure 
of the wind (i.e. density and size) can change as a function of
the luminosity.
Similarly observed variations in Cir~X-1 have been interpreted to be due to changes
in the ionization fraction of the wind (Schulz \& Brandt, 2001, in preparation). 
We note however that the statistics are not adequate for us  to present a
complete quantitative picture for \grs1915, especially in light of the 
numerous problems we have with pileup that need to be considered.  
Cycle~2 \chandra data of
\grs1915 taken in a mode that optimally reduces pileup will be able to
study the variability aspects in greater detail.

\section{Discussion}
The \chandra HETGS spectrum of \grs1915 shows neutral and ionized
features superposed on a continuum best described by a disk blackbody
and a dominating power-law modified by a column of $\sim 5 \times 10^{22} \rm \,
cm^{-2}$ absorption.
The X-ray spectral and temporal characteristics as well as the 3.7~Hz
QPO seen in our data, and the detection of faint radio emission are
all consistent with the low$-$hard state. This indicates that the
X-ray power-law is the dominant emission component and there is a weak
and persistent radio jet.  No obvious red- / blue-shifted features
indicative of a jet can be seen in the  time-averaged \chandra
spectrum.

The H column density derived from the Si and Fe edges (assuming solar
abundances) is  $\sim 8-9 \times 10^{22} \psqcm$, which is substantially
higher than the values derived from S and Mg, and continuum absorption
determined from broad-band instruments.   The discrepancy for excess Si 
and Fe absorption is even higher if we assume ISM abundances. 

There is a possibility that the high X-ray column density and 
abundance excesses may be related to 
material that is  associated with the immediate environment of the
source.  Both Mirabel et al. (1996) and Mart\'{\i} et al. (2000)
have suggested the possibility for heated dust that is 
associated with the surroundings of \grs1915 from IR observations.
If true, these anomalous Si and
Fe  abundances may  provide hints to the evolutionary events
which gave rise  to this source, and the black hole in particular.
However, we acknowledge that more data from other X-ray binaries
with similar galactic column densities are necessary to determine 
the absorption abundance ratios  over the Galaxy and so confirm 
that the results for \grs1915 are exceptional.
Thus far, observations of SS~433 (Marshall et al. 2001) and Cyg~X-1 
(Schulz et al. 2001) do not show such Fe and Si overabundances.
(See also e.g. respectively Paerels et al. 2001 and Schulz et al. 2001 
for the X-ray binaries X0614+091 and Cyg~X-1, and Lee et al. 2001b
for the extragalactic source \mcg6.)
Clearly, further sensitive
measurements such as these of photoelectric edges with \chandra will provide
valuable information for assessing the material which surrounds bright
X-ray sources especially if depletion plays a prominent role.
Such measurements will be particularly important for confirming 
the tentative detection of XAFS (particularly in Si) as reported here.  
If true, such a positive detection of these absorption fine structure 
representative of material in solids (as e.g. interstellar grains) will 
have important consequences for conducting solid state astrophysics, by 
which grain properties can be extracted via the solid's inner compound 
structure. 

In the interim, the abundance excesses for X-ray absorption edges in
GRS~1915+105 can be contrasted with the discovery of abundance
anomalies in the optical spectra of companion stars in the
microquasars GRO J1655-40 and V4641 Sgr (Israelian et al. 1999; Brown
et al. 2000; Orosz et al. 2001).  In the latter cases, the
over-abundances of $\alpha$-process elements, but not the Fe group,
has led to the hypothesis that the companion star captured supernova
or hypernova ejecta is related to the formation of the black hole.  The
Fe in the progenitor's core is likely to be efficiently captured by
the black hole. In the case of GRS 1915+105, the Fe and Si
overabundance pertains to material along the line of sight, and the
enrichment may signify either a highly unusual supernova or perhaps
external supernova activity local to the binary.  

One way to reconcile the overabundance of Fe in \grs1915 is if an
asymmetric  hypernovae/supernova explosion produced heavier elements
like Fe and Si along the poles.   For the $14 \Msun$ black hole in
\grs1915, the binary spin axis would then have to be misaligned with the
$70^{\circ}$ inclination of the jet (Mirabel \& Rodr\'iguez 1994) with
respect to the line of sight. Such a misalignment could be induced by
disk warping, precession of the spin axes, or residual angular
momentum from the initial supernova explosion.

One rather speculative possibility is that the
excess Fe and Si might indicate a connection between
some kind of SN and BH microquasars. 
Given that BH microquasars are likely formed from SN, Gamma-ray bursts (GRB)
may also represent some kind of microquasar 
(e.g. Blackman et al. 1996; Kouveliotou et al. 1999)
formed similarly (Paczynski 1998; Brown et al. 2000). 
This makes contact with the results of Brown et al. (2000)
for GRO 1655+40, although as mentioned above, there the overabundant
elements are alpha-process elements S and O which are more
easily understandable as a SN signature. Nevertheless,
one might ultimately imagine a classification of microquasars 
analogous to that of AGN. 

In addition to the first detection of photoelectric edges in \grs1915,
our \chandra observation confirms the presence of
 absorption features from highly ionized gas previously reported by 
Kotani et al. (2000) with \asca data taken between in 1994 to 1999.
Despite their use of a very complicated continuum model, it is
evident from Fig. 1 of Kotani et al. (2000) that the observations
showing strong absorption features (years 1994 and 1995) display steep X-ray
spectra near 10 keV, in contrast with the 1996 observation, where the
spectrum is harder, and no absorption features were found with the \asca data.
 We confirm with the \rxte PCA data archive that GRS1915+105 was in a
bright mode (1.2 Crab) of the hard-steady state on 1996 October 23
(see Muno et al. 2001). One might therefore suspect from the reported \asca
observations that the absorption features from highly ionized atoms
are confined to the soft X-ray states of GRS~1915+105; however, the
\chandra results shown here clearly show such absorption features in
the low-hard state (0.4 Crab), as well. This is consistent with the 
view that the absorbing gas is associated with the accretion disk,
which appears evident as the soft component in the X-ray spectrum.

Given that \grs1915 is probably accreting near or above its Eddington
limit, the presence of a strong radiatively-driven wind is almost
expected.  However, we do not find obvious P-Cygni profiles attributed to a
disk wind --- i.e., redshifted/rest-frame emission lines (from
material out of the line of sight) accompanied by blueshifted
absorption lines (from the foreground, line-of-sight parts of the
wind) in the time-averaged data, akin to what has been reported for Cir~X-1 (Brandt \& Schulz
2000).  The lack of  
a wind from the star may corroborate the finding that the donor in \grs1915 is a
LMXB (Greiner et al., 2001a) where accretion via Roche lobe overflow is
the most plausible mechanism for mass transfer.  However, the wind (and/or jet
signatures) if present (e.g. \S\ref{sec-fe}) in \grs1915 is likely to originate from the disk, 
and may show up in a more subtle form (e.g. through variability studies, \S\ref{sec-var}).
The ionized Fe absorption lines of the time averaged  spectrum are
likely to have origins from the disk or disk atmosphere.

This hot absorbing medium is a particularly important topic, since it
points to a disk atmosphere, and possibly even disk wind, or some
other type of hot gas structure that is absent from most models for
black hole accretion.  It will be interesting to diagnose the effects
of scattering  and absorption in this medium with \chandra spectra of
\grs1915 in different flux states, in order to build a picture of which
spectral changes are  inherent to the source and which are tied to
changes along the line of sight to the source. 
The latter topic is a
highly controversial one for \grs1915.  The violent changes seen in
the apparent disk radius and temperature have been interpreted on some
occasions as physical changes that represent the disk instability
(Belloni et al 1997) and the effects of the jet production mechanism
(Eikenberry et al. 1998).  However, analogous disk changes,
particularly when the variations are more modest, have been
alternatively interpreted as effects due to radiative transfer in an
accretion disk (Merloni Fabian \& Ross 2000).  
The \chandra spectra (\S\ref{sec-var}) suggest that
the apparent changes in the strength of the Fe~{\sc xxv} and Fe~{\sc xxvi} 
absorption can be linked to changes in the incident ionizing flux and density,
and may point to a flow which
can change over $\sim 10$~ks time scales.
Clearly, longer continuous observations with \chandra and \rxte are needed for
this source in order to monitor the observed variability, and differentiate 
between the different mechanisms which may be responsible.

\section*{acknowledgments}
We wish to thank John Houck for assistance with {\sc isis}, Claude 
Canizares for critical discussions and useful suggestions, and
John Davis for expert advice on pileup correction.  We also thank
Thomas Dame for advice on the interpretation of line of sight column
densities, Tim Kallman, Frits Paerels, J\"orn Wilms, Anil Pradhan
 and Justin Oelgoetz for advice.  We also thank
many of our colleagues in the MIT HETG/CXC group, and 
acknowledge the great efforts of the
many people who contributed to the \chandra program.  
This work was funded by the Chandra grant GO0-1103X.   JCL also acknowledge
support from the NASA contract NAS8-39073, and NSS from SAO SV1-61010.
CSR acknowledges support from Hubble Fellowship grant HF-01113.01-98A --
this grant was awarded by the Space Telescope Institute, which is operated
by the Association of Universities for Research in Astronomy, Inc., for
NASA under contract NAS 5-26555.
ACF thanks the Royal Society for support.

\end{document}